\documentstyle[a4,12pt,epsf]{article}

\begin{document}

\title{
{\large\hfill NBI preprint Nr.\ 96--39}\\[1em]
Hyperon--rich Matter and Kaons in Matter
\footnote{Talk given at the International Conference on Strangeness
in Hadronic Matter, May, 15--17 1996, Budapest, Hungary, to be
published in Heavy Ion Physics}
}

\author{
J\"urgen Schaffner$^1$, Jakob Bondorf$^1$ and Igor
N. Mishustin$^{1,2}$\\[1em]
\normalsize
$^1$The Niels Bohr Institute, Blegdamsvej 17, DK-2100 Copenhagen \\
\normalsize
$^2$The Kurchatov Institute Russian Research Center, Moscow 123182 
}

\date{}

\maketitle

\begin{abstract}
We discuss the equation of state of neutron stars in the dense
interior considering hyperons and the possible onset of kaon
condensation within the relativistic mean field model. We find
that hyperons are favoured in dense matter and that their appearance
make the existence of a kaon condensed phase quite unlikely.
Implications for the recent measurements of kaons in heavy ion
collisions at subthreshold energy are also given.
\end{abstract}

\section{Introduction}

Nuclear matter at high densities and temperatures
exhibits a new degree of freedom:
strangeness, in neutron star matter
hyperons and possibly kaons appear at a moderate density of about
$2\div 3$ times normal nuclear matter density $\rho_0=0.15$ fm$^{-3}$.
These new species influence the properties of the equation of state
of matter and the global properties of neutron stars.

The Relativistic Mean Field (RMF) model has been used first by Glendenning
for describing the equation of state of matter with hyperons 
\cite{Glen87}. He founds that hyperons appear at 
$\rho\approx (2-3)\rho_0$ rather independent of the equation of state (EOS)
used. Hyperons considerably soften the EOS and reduce the maximum mass
of a neutron star.
The appearance of kaons was discarded, as the chemical potential was
saturating around 200 MeV while the kaon mass stayed constant with
density, i.e.\ in-medium effects (as a decrease of the kaon energy)  
were not taken into account.

On the other hand, chiral perturbation theory (ChPT) gives a rather robust
prediction of the onset of antikaon condensation at 
$\rho\approx (3-4)\rho_0$ \cite{Brown} taking into account 
in-medium modifications of the antikaon energy with density. 
Antikaon condensation will soften the EOS and reduce the maximum mass
of a neutron star similar to the case of hyperons. 
This will allow for the existence of low-mass black
holes and implications have been discussed in \cite{Bethe}.
In \cite{Maru94,Sch94} it was
criticised that effects nonlinear in density were not taken into
account which will shift the appearance of a kaon condensed phase to
higher density. Moreover, hyperons were not considered in this approach.  

Rather recently, antikaons and hyperons were considered on the same
footing, where it was found that hyperons shift
the onset of antikaon condensation to higher density
\cite{Prakash} or that it is very unlikely that
it appears at all due to the rather strong hyperon-hyperon
interactions \cite{Sch96}. 
In the following we outline our recent work \cite{Sch96}
within the RMF model in more detail. 
Furthermore, we will discuss implications of
in-medium effects of kaons and antikaons for the recent heavy ion
experiments at subthreshold energy.

\section{Strangeness in Neutron Stars}

\subsection{The RMF Model with Hyperons}

The implementation of hyperons within the RMF approach is straightforward.
SU(6)-symmetry is used for the vector coupling constants
and the scalar coupling constants are fixed to the potential depth of the
corresponding hyperon in normal nuclear matter \cite{Sch93}.
We choose
\begin{equation}
U_\Lambda^{(N)} = U_\Sigma^{(N)} = -30 \mbox{ MeV} \quad , \qquad
U_\Xi^{(N)} = -28 \mbox{ MeV} \quad.
\label{eq:potdep1}
\end{equation}
Note that a recent analysis \cite{Mar95}
comes to the conclusion that the $\Sigma$ potential changes 
sign in the nuclear interior, i.e.\ being repulsive instead of attractive.
In this case, $\Sigma$ hyperons will not appear at all in our
calculations. Nevertheless, our conclusion about the onset of kaon
condensation remains unchanged.

The observed strongly attractive $\Lambda\Lambda$ interaction
is introduced by two additional meson fields, the scalar meson $f_0(975)$
and the vector meson $\phi(1020)$.
The vector coupling constants to the $\phi$-field are given by SU(6)-symmetry
and the scalar coupling constants to the
$\sigma^*$-field are fixed by 
\begin{equation}
U^{(\Xi)}_\Xi \approx U^{(\Xi)}_\Lambda \approx
2U^{(\Lambda)}_\Xi \approx 2U^{(\Lambda)}_\Lambda \approx -40 \mbox{ MeV}
\quad .
\label{eq:potdep2}
\end{equation}
Note that the nucleons are not coupled to these new fields.

\subsection{Neutron Stars with Hyperons}

Fig.~1 shows the composition of neutron star matter 
for the parameter set TM1 with
hyperons including the hyperon-hyperon interactions.

Up to the maximum density considered here all effective masses
remain positive and no instability occurs. 
The proton fraction
has a plateau at $(2-4)\rho_0$ and exceeds 11\%{} which allows for the
direct URCA process and a rapid cooling of a neutron star.
Hyperons, first $\Lambda$'s and $\Sigma^-$'s, appear at $2\rho_0$, then
$\Xi^-$'s are populated already at $3\rho_0$. The number of electrons
and muons has a maximum here and decreases at higher densities, i.e.\
the electrochemical potential decreases at high densities. 
The fractions of all baryons show a tendency towards
saturation, they asymptotically reach similar values 
corresponding to spin-isospin and hypercharge-saturated matter.
Hence, a neutron star is more likely a giant hypernucleus \cite{Glen87}!

\begin{figure}
\epsfysize=0.7\textwidth
\vskip -0.5cm
\centerline{\epsfbox{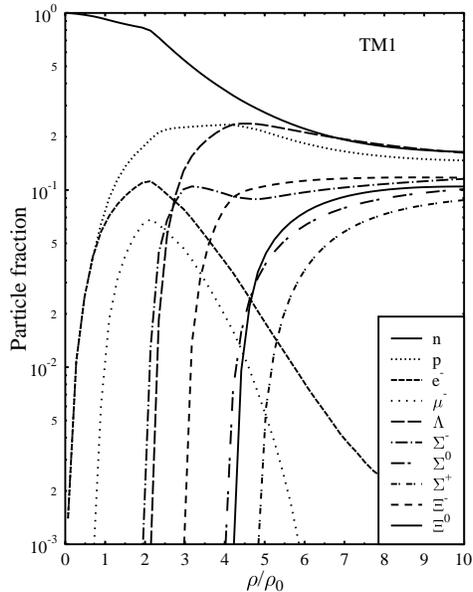}}
\vskip -1cm
\caption{
The composition of neutron star matter with hyperons which appear
abundantly in the dense interior.}
\end{figure}

\subsection{Kaon Condensation ?}

In the following
we adopt the meson-exchange picture for the KN-interaction simply
because we use it also for parametrizing the baryon interactions.
We start from the following Lagrangian
\begin{equation}
{\cal L}^{\rm RMF}_{KN} = D^*_\mu \bar K D^\mu K - m_K^2 \bar K K
- g_{\sigma K} m_K \bar{K}K \sigma
- g_{\sigma^* K} m_K \bar{K}K \sigma^*
\label{RMFLagr}
\end{equation}
with the covariant derivative
\begin{equation}
D_\mu = \partial_\mu +
ig_{\omega K} V_\mu + ig_{\rho K} \vec{\tau}\vec{R}_\mu + ig_{\phi K}\phi_\mu
\quad .
\end{equation}
The scalar fields essentially decrease the kaon mass, while
the vector fields will increase (decrease) the energy of the kaon
(antikaon) in the dense medium.
The scalar coupling constants are fixed by the s-wave KN-scattering lengths.
The coupling constants to the vector mesons are chosen from
SU(3)-relations.
The onset of s-wave kaon condensation is now determined by the condition
$- \mu_e = \mu_{K^-} \equiv \omega_{K^-} (k=0)$.

The density dependence of the K and $\bar K$ effective energies is displayed
in Fig.~2.
The energy of the kaon is first increasing in accordance with 
the low density theorem.
The energy of the antikaon is decreasing steadily at low densities.
With the appearance of hyperons the situation changes dramatically.
The potential induced by the $\phi$-field cancels the contribution
coming from the $\omega$-meson. Hence, at a certain density the energies of
the kaons and antikaons become equal to the kaon (antikaon) effective mass, 
i.e.\ the curves for kaons and antikaons are crossing at a 
sufficiently high density. At higher densities
the energy of the kaon gets even lower than that of the antikaon!
Since the electrochemical potential never reaches values above 160 MeV here 
antikaon condensation does not occur at all.
We have checked the possibility of antikaon condensation for all parameter
sets and found that at least 100 MeV are missing
for the onset of kaon condensation. This is in contrast to previous 
calculations disregarding hyperons \cite{Brown} but in line with 
the findings in \cite{Prakash} where it was seen that hyperons shift
the critical density for kaon condensation to higher density.

\begin{figure}
\epsfysize=0.65\textwidth
\vskip -0.5cm
\centerline{\epsfbox{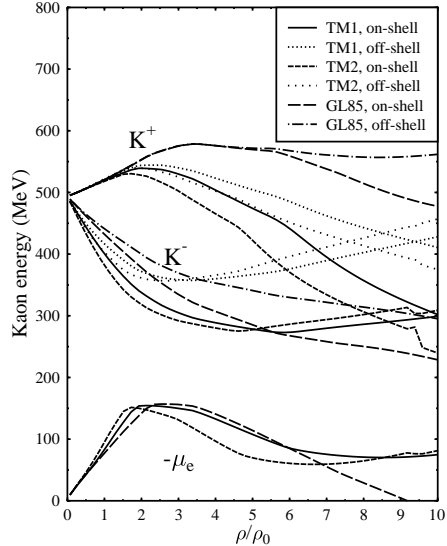}}
\vskip -1cm
\caption{
The effective energy of the kaon and the antikaon.
and the electrochemical potential.
Kaon condensation does not occur over the whole density region
considered.}
\end{figure}

\section{Kaons in Heavy Ion Collisions}

Subthreshold production rates of K$^+$ in heavy-ion
reactions were recently measured at GSI \cite{GSI}.
The influence of rescattering and formation of resonance 
($\Delta$) matter was studied in the QMD model \cite{Hart94},
and the RBUU model \cite{Maru94b} and it was demonstrated that they
are essential to explain the data.
In-medium modifications
of the effective energy of the kaon were studied in \cite{Fang94}
using again the RBUU model. The results are essential similar
to the ones obtained without medium modifications \cite{Maru94b},
because the in-medium kaon mass
used is quite close to the respective vacuum mass.
But there exist other observables which might be better suited
for extracting in-medium effects:
the flow of kaons might be a promising tool for
measuring the kaon potential in dense matter \cite{kaonflow}.
And more pronounced in-medium effects are expected for the case of
K$^-$ \cite{Li94}. Indeed, enhanced production rates for K$^-$
have been seen at GSI recently \cite{GSI2}.

In the following we discuss the in-medium change of the threshold
energy for kaon production. Details will be found in a forthcoming
publication \cite{Schnew}.

\subsection{RMF model for Kaons}

The Lagrangian has been given above (eq.\ \ref{RMFLagr}). From this
one gets the following kaon self energy in the mean field approximation
\begin{equation}
\Pi (\omega,k;\rho_N) = 
- g_{\sigma K}\sigma m_K 
+ 2 g_{\omega K} \omega V_0 + (g_{\omega K} V_0)^2 
\quad .
\end{equation}
Note that the scalar and vector fields depend on the nucleon 
density $\rho_N$ and that the kaon self energy depends on the kaon
energy itself.
The antikaon energy is then given by
\begin{equation}
\omega_{\bar K}(k) =
\sqrt{m_K^2 + m_K g_{\sigma K} \sigma + k^2 } - g_{\omega K} V_0
\end{equation}
and the coupling constants are fixed again to KN scattering lengths
\cite{Sch96}.

\subsection{Chiral Perturbation Theory for Kaons}

We follow the procedure outlined in \cite{Brown}
starting from the Nonlinear Chiral Lagrangian in next-to-leading order
\begin{equation}
{\cal L}^{\rm ChPT}_{KN} =
- \frac{3i}{8f_K^2}\left[
\bar{N}\gamma_\mu N \left(\bar{K} 
\stackrel{\leftrightarrow}{\partial^\mu} K \right) \right]
+\frac{\Sigma_{KN}}{f_K^2}\bar{N}N\bar{K}K
+\frac{\tilde{D}}{f_K^2}\bar{N}N 
 \left(\partial_\mu\bar{K}\partial^\mu K \right)
\end{equation}
where $f_K=93$ MeV is the kaon decay constant and
$\Sigma_{KN}=2m_\pi$ is the sigma term.
The first two terms comes from the Tomozawa--Weinberg term and is in
leading order of the chiral expansion. This is a vector interaction
and is repulsive (attractive) for kaons (antikaons). 
The other terms are in next-to-leading order. The second term involves
the sigma term which is of scalar type
and will decrease the mass of the
kaon and antikaon in the medium. The last term
is the so called off-shell term which will modify the scalar
attraction. It is essential to get a correct scattering length \cite{Brown}.
The kaon self energy $\Pi (\omega,k;\rho_N)$ is given in the mean
field approximation by
\begin{equation}
\Pi (\omega,k;\rho_N) = 
- \frac{\Sigma_{KN}}{f_K^2}\rho_s
- \frac{\tilde{D}}{f_K^2}\rho_s \omega^2
+ \frac{3}{4f_K^2} \omega \rho_N 
\end{equation}
which depends in general also on the kaon energy. 
This has to be taken into account
to get the energy of a kaon or antikaon in the nuclear medium correctly.

\begin{figure}
\epsfysize=0.45\textheight
\centerline{\epsfbox{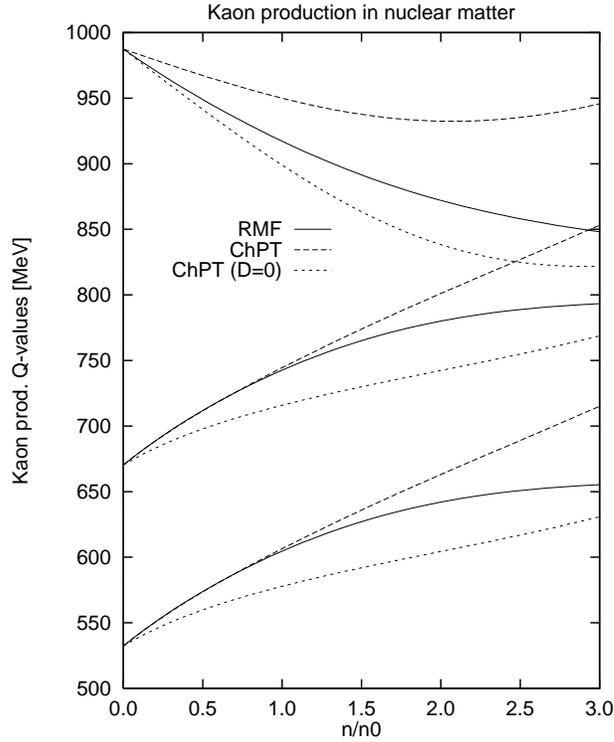}}
\caption{
The threshold energy of the processes 
NN$\to$N$\Lambda$K (middle curves), NN$\to$NNK$\bar K$ (upper curves)
and $\pi$N$\to\Lambda$K (lower curves) versus the density.}
\end{figure}

\subsection{Subthreshold Production of Kaons}

In the following we discuss the shift of the threshold energy
of various processes for heavy ion collisions due to medium modifications.

Kaons are mainly produced at threshold via the process
NN$\to$N$\Lambda$K. The minimum energy needed is 
$Q({\rm N}\Lambda{\rm K})\approx 671$ MeV in vacuum.
In the medium, the threshold is shifted to
\begin{equation}
Q({\rm N}\Lambda{\rm K})=E_\Lambda(p=0) + \omega_K(k=0) - E_N(p=0)
\end{equation}
where we assume that the outgoing nucleon is not Pauli-blocked in 
the hot zone of the collision. Hence, the subthreshold production
of kaons is sensitive to three different in-medium effects:
the EOS ($E_N$), the $\Lambda$ potential ($E_\Lambda$) and the
kaon energy ($\omega_K$) in medium. 
These effects will partly cancel each other as the kaon feels a
repulsive potential of $29$ MeV due to the low density theorem while
the $\Lambda$ sees an attractive potential of $-30$ MeV
(see e.g.\ \cite{Sch93}). Therefore, subthreshold kaon production 
seems to probe mainly the nuclear EOS. As the nucleons feel an attractive
potential of about $60$ MeV the threshold will be shifted {\em
upwards} at normal nuclear density by this amount and the production
of kaons is {\em reduced} in the medium. This is indeed the case
as can be seen from the middle curves of Fig.~3 which shows the
threshold energy $Q({\rm N}\Lambda{\rm K})$ as a function of density.
The similar behaviour of the different curves at low density is due
to the low-density theorem. 
At $\rho=3\rho_0$ the value of $Q({\rm N}\Lambda{\rm K})$ reaches
about 800 MeV for the RMF model and about 860 MeV for ChPT.
Without off-shell terms, the threshold
energy is underestimated in medium, and we expect that the production
rates for kaons calculated in \cite{Fang94} are overestimated.
Note, that all calculations ignoring in-medium effects
\cite{Hart94,Maru94b} will also
give a too high production rate for kaons.
Note that the threshold energy for the 
secondary process $\pi$N$\to\Lambda$K will be also
shifted upwards with density (lower curves in Fig.~3).

Antikaons are created in heavy ion collisions first by
the process NN$\to$NNK$\bar K$. The threshold value of 
$Q({\rm K}\bar K)\approx 988$ MeV is modified in the medium and given by 
the sum of the kaon and antikaon energy 
$Q({\rm K}\bar K)=\omega_K(k=0) + \omega_{\bar K}(k=0)$.
Therefore, subthreshold antikaon production probes the in-medium
property of kaons and antikaons solely. As the vector potential
cancels out approximately, it will mainly depend on the scalar potential
the kaon feels in the medium. The upper curves in Fig.~3
show that indeed 
$Q({\rm K}\bar K)$ is reduced in the medium in all models discussed
here. ChPT predicts an in-medium reduction of about $-56$ MeV at
maximum compared to the vacuum and then the curves go up again
for higher density.
On the other side, the RMF model gives a reduction of about
$-140$ MeV at $\rho=3\rho_0$. The $Q$-value is steadily decreasing.
The curve used in RBUU calculations neglecting off-shell terms
\cite{Li94} is lying even lower
and hence, the in-medium production rates of antikaons seems to be
overestimated.

As an interesting fact, the $Q$-values for
kaon and antikaon production are lying close together for the RMF
model. Note that this does not mean that the numbers of produced
kaons and antikaons are the same. 
The kaons will be produced at different density and the average
$Q$-value over the density profile will give a measure for the total
number of produced
kaons and antikaons in the medium.

\section{Conclusions}
 
We discussed medium modifications of the kaon and antikaon on the mean
field level and applied it for neutron star matter and heavy ion reactions.
We have shown that kaon condensation is unlikely to happen inside a
neutron star within the RMF model due to the presence of hyperons. 
Relativistic heavy ion reactions at subthreshold energy
might see in-medium modifications of the kaon or antikaon.
The threshold energy for kaon production goes up with density.
Hence, rescattering effects and resonance decays are mainly
responsible for the enhanced K$^+$ production seen. 
On the other hand, the antikaon production is enhanced in dense matter
due to a decreased antikaon energy. This attractive potential will be
visible in e.g.\ the antikaon flow or the excitation function of the
produced K$^-$.

\section*{Acknowledgements}

We like to thank the organisers of the conference for their support and
Avraham Gal, Norman K. Glendenning, 
Vesteinn Thorsson, Fridolin Weber,
Wolfram Weise and Thomas Waas for useful discussions and remarks.
J.S. thanks Che Ming Ko, Madappa Prakash, 
Peter Senger and Andreas Wagner for enlightening conversations during 
this conference.
I.N.M. is supported in part by 
the Carlsberg Foundation (Denmark), 
the International Science Foundation (Soros)
grant N8Z000 and EU-INTAS grant 94-3405.
Two of us (J.S. and I.N.M.) thank the Niels Bohr Institute for kind
hospitality and financial support.

\end{document}